# Modelling Electrical Car Diffusion Based on Agent

[1]Lei Yu, ,[2]Tao Zhang, [3]Siebers Peer-Ola, [4]Uwe Aickelin
*1 School of Electronic and Control, Chang'an University, Xi'an, china, 710064.
Yul525@163.com
2,3,4 School of Computer Science, the University of Nottingham, Nottingham, UK, NG8 1BB

### Abstract

Replacing traditional fossil fuel vehicles with innovative zero-emission vehicles for the transport in cities is one of the major tactics to achieve the UK government 2020 target of cutting emission. We are developing an agent-based simulation model to study the possible impact of different governmental interventions on the diffusion of such vehicles. Options that could be studied with our what-if analysis to include things like car parking charges, price of electrical car, energy awareness and word of mouth. In this paper we present a first case study related to the introduction of a new car park charging scheme at the University of Nottingham. We have developed an agent based model to simulate the impact of different car parking rates and other incentives on the uptake of electrical cars. The goal of this case study is to demonstrate the usefulness of agent-based modelling and simulation for such investigations.

## 1. Introduction

### 1.1 Background

Energy consumption in the transport sector is the major part of city energy consumption. According to DECC (The Department of Energy and Climate Change), in 2009 77% of total oil consumption in the UK was consumed in the transport sector [1], Fig 1. Burning fossil fuels such as petroleum, diesel and gas to drive vehicles often causes huge $CO_2$ emission. Thus replacing traditional fossil fuel vehicles with innovative zero-emission vehicles for the transport sector in cities is one of the major tactics to achieve the UK government's 2020 target of cutting emission.

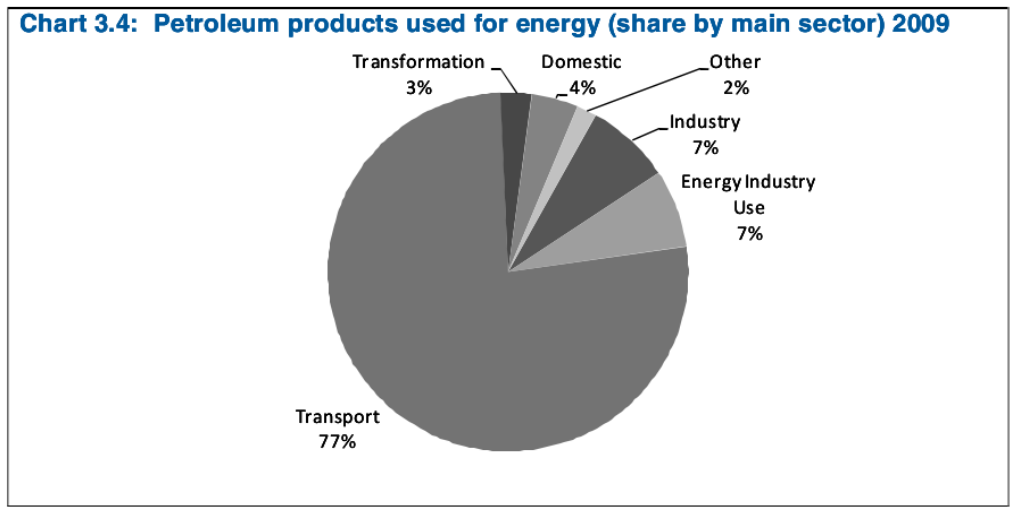

**Figure 1.** city energy consumption in UK, 2009

In order to improve people to buy electrical car, in July, 2010 UK government announced a project that anyone who buys an electric plug-in car from 2011 will get a 25% discount up to a maximum of £5,000 [2]. It means that anyone who buys an electric plug-in car from next year will get a 25% discount up to a maximum of £5,000.

In the paper we focused on the diffusion of electric vehicles in cities. Essentially, we look at the policy of differentiating parking charge rates for promoting electric vehicle diffusion. In the UK, as parking charge rates in different cities are managed by local authorities, differentiating parking charge rates for different types of vehicles is one of the options that could be taken by local authorities. However, the effectiveness of such a policy is still uncertain. This paper aims to provide an comprehensive study to evaluate the effectiveness of this policy.

## 1.2    Agent-based simulation

Theoretically speaking, agents are the constituent units of a multi-agent system (MAS). Agents are autonomous, behave on their own, and interact with each other in an MAS [3]. The behavior and interactions of agents produce the global behavior of an MAS. However, this type of global behavior of an MAS cannot be conversely traced back to the behavior and interactions of its constituent agents. Agents can be objects that are of low or medium level intelligence, (e.g. machines and software programs, which just have some functional, procedural or algorithmic search, find and processing approach), or objects that are highly intelligent (e.g. human beings and societies) [4]. Agent-based simulation is a computational modeling approach to study MASs. An agent-based model is composed of individual agents, commonly implemented in software as objects. Agent objects have states and rules of behavior. Running an agent-based model simply amounts to instantiating an agent population, letting the agents behave and interact, and observing what happens globally [5]. Thus a unique advantage of agent-based simulation is that almost all behavioral attributes of agents can be captured and modeled. Agent-based simulation is widely adopted in studying MASs, particularly those with intelligent human beings (e.g. markets, societies, and organizations; for further information about agent-based simulation and its applications [6].

## 1.3    Related Work

Traditionally, in energy research policies are assessed by static econometrics models, e.g. input-output analysis and cost-benefit analysis. However, in recent years, agent-based modelling (ABM) and simulation has received increasing attention from many researchers in the field of energy policy assessment and energy system modeling[7, 8, 9, 10]. For example, Stephan and Sullivan [11] put forward an agent based model to study the transition of a transport system based on conventional fuels to one based on alternative fuel, such as hydrogen. Zhang and Levinson [12] applied the method to form an agent-based travel demand model for predicting macroscopic travel patterns from microscopic agent travel behaviours. Han et al [13] developed an ABM to characterize macro-level time based vehicle distributions based on agent choice of trip departure times, which change with agent traffic experience.

Ting Zhang et al [14] developed an ABM model to investigate factors that can speed the diffusion of eco-innovations. In the model he considered three factors to influence the alternative fuel vehicles diffusion: technology push, market push and regulatory push. For the electrical car diffusion, it doesn't follow the typical Bass diffusion curve because of long times[15] and diffusion discontinuities[16]. Slow diffusion is a common occurrence with eco-innovation.

## 2.    Model Design

## 2.1    Main concepts for the Model

Fig2 shows the main concepts for the model. In the model there have two different type agents. One is car owner, the other is parking place. The car owner agents parks the car in parking place every work day and goes back to home (pool of car owner agents) after work. The electrical car diffusion is through Word of Mouth or Ads. Each car owner has many properties, such as staff level, type of vehicles, energy awareness, etc. The parking place has two properties: free and occupied.

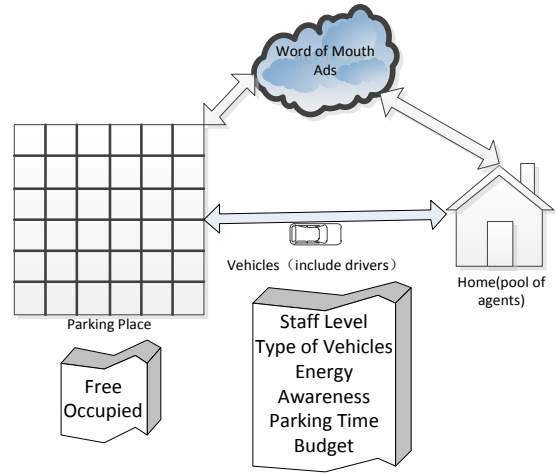

**Figure 2.** Main concepts for the model

## 2.2 Concepts for the agents

We have used state charts for the concepts of the agents. Fig 3 and Fig 4 shows two types of agent in the model. One is car owner agent, the other one is parking place agent.

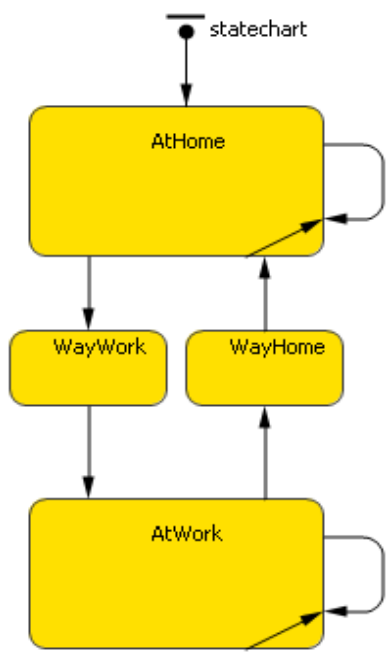

**Figure 3.** Car Owner Agent

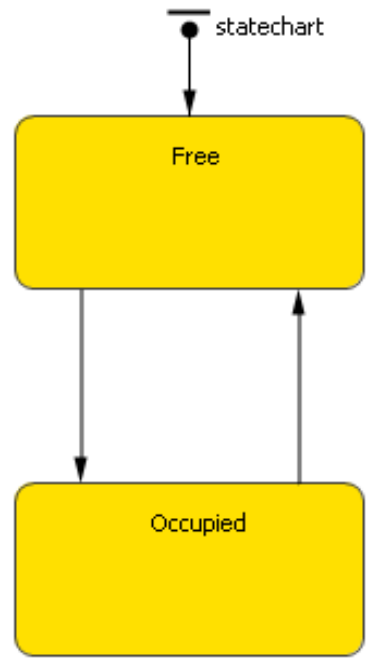

**Figure 4.** Parking Place Agent

The model of the car owner has four main states: AtHome, AtWork, WayHome, WayWork. The model of the parking place has two states: Free, Occupied.

When the car owner is in "AtHome" state, make the agent invisible. When the car owner begins to work, he should park the car in a free space. So he should send a message to the parking place agent, When there has a free place, the parking place agent becomes to "Occupied" state ( filling colour becomes red )  and the car owner parks the car in this place( make the car visible). When the car owner agent goes home, this parking place agent changes from occupied to free (filling colour becomes white).

From Fig 3 we can see that in the "AtHome" state and in the "AtWork" state there also internal transition. This transition models how this car owner persuades his acquaintance to purchase the electrical car. Transition's rate depends on this person's cogency and contact rate. WOM transition should be taken when the state chart of the agent receives the "Buy Electrical Car" message sent by its acquaintance. Ads transition means this car owner can be influenced by ads.

## 2.3 Model Validation

I used two steps to validate this model. The first step is face validation. I introduce my model to the estate office of university. The second step is comparing this model to another system dynamic model of WOM[8]. We found they have close results.

## 3. Implementation

Our model has been implemented in anylogic 6.6.0. Anylogic is a Java based simulation software. In this part we will introduce the implementation of the model, includes Car Parking Charge Policy in Jubilee Campus and definition for each type of car owner agents.

## 3.1 Car Parking Charge Policy in Jubilee Campus

Currently, In Jubilee campus the car parking charges is designed as table 1. We can see that there haven't different charges between traditional car and electrical car. Our first experiment focuses on this and aims to design different car parking charge strategies.

**Table 1.** Car Parking Charge Policy in Jubilee Campus

| Category | Emissions Category | Staff Level 1 | Staff Level 2 | Staff Level 3 | Staff Level 4 | Staff Level 5 | Staff Level 6 | Staff Level 7 |
|----------|-------------------|---------------|---------------|---------------|---------------|---------------|---------------|---------------|
| A | Up to 120 g$CO_2$/km | 44 | 57 | 75 | 105 | 132 | 165 | 210 |
| B | 121-150 g$CO_2$/km | 58 | 76 | 100 | 140 | 176 | 220 | 280 |
| C | 151-165 g$CO_2$/km | 73 | 95 | 125 | 175 | 220 | 275 | 350 |
| D | 166-200 g$CO_2$/km | 87 | 114 | 150 | 210 | 264 | 330 | 420 |
| E | >200 g$CO_2$/km | 102 | 133 | 175 | 245 | 308 | 385 | 490 |

## 3.2 Definitions for each type of staff

For the car owner agent, we defined four types, we assume that each person has a personality parameter energy awareness, ranging from 0 to 100, to represent its awareness on energy saving. If a person's energy awareness is greater than a threshold, it has a large probability to buy the electrical car. In the simulation, the threshold is adjustable, with value ranging from 0 to 100. Based on our empirical survey on staff's energy awareness, we create four stereotypes of user agent for the simulation model. Different stereotypes of user agents have different levels of energy awareness, and the probabilities for them to buy electrical car, as shown in Table 2.

**Table 2**. Stereotypes of Car Owner Agents

| Stereotype of Agent | Percentage | Energy Awareness | Probability of Buying Electrical Car |
|---|---|---|---|
| 1 | 1% | Between 95 and 100,random uniform distribution | 0.9 |
| 2 | 9% | Between 70 and 94,random uniform distribution | 0.7 |
| 3 | 30% | Between 30 and 69,random uniform distribution | 0.4 |
| 4 | 60% | Between 0 and 29,random uniform distribution | 0.2 |

## 4.    Experiments

Using this model, I developed two experiments. I use these experiments to design different car parking charge strategies and help the diffusion of electrical car.

### 4.1    Experiment 2: Comparing Different Car Parking Charges

Now in Jubilee Campus there haven't different car parking charges between traditional car and electrical car, the staffs haven't motivation to change their traditional car. In this experiment I focus on the car parking charge strategy. The initial parameters are set up as follow: number of car parking place is 600, number of staff is 500, and energy awareness threshold is 50. Fig 5 shows that electrical car number distribution, fig 6 shows the whole car parking charges changes, and fig 7 shows the energy consumption distribution. We can see that decreasing electrical car parking charges can help the diffusion of electrical car; the whole energy consumption will decrease accordingly.

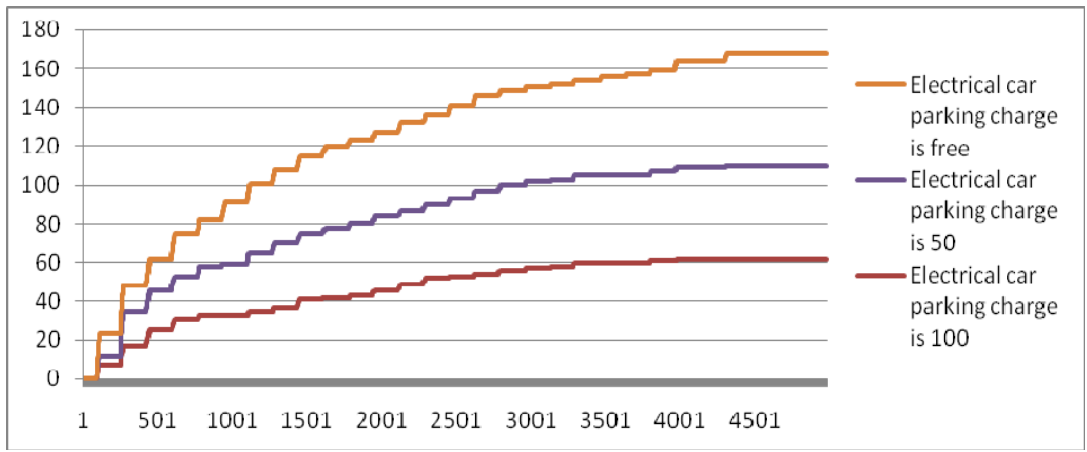

**Figure 5.**  electrical car number distribution

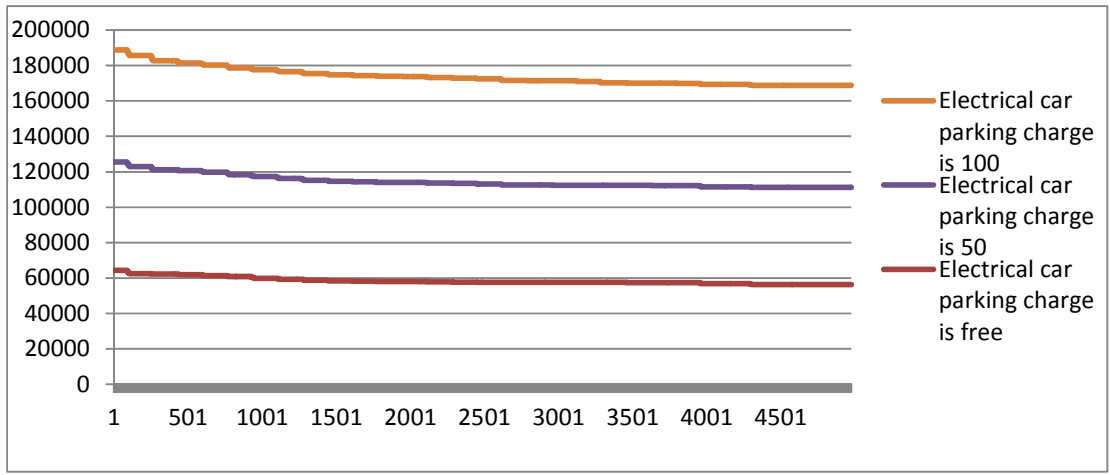

**Figure 6.** Whole car parking charges changes

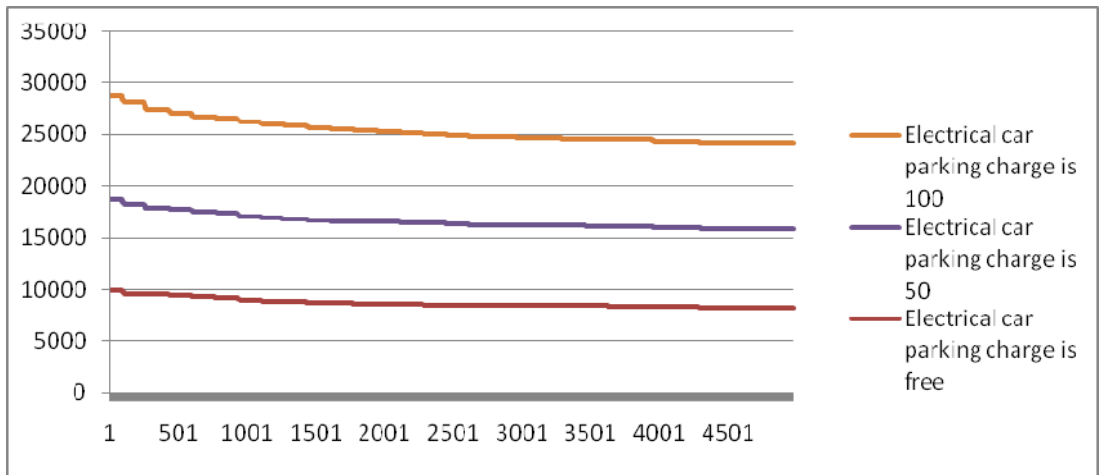

**Figure 7.** Energy consumption distribution

### 4.2    Experiment 2:Consideration of WOM

In this experiment we will discuss the factors of word of mouth. Each car owner agent will be influenced by Ads, WOM (Word of Mouth) and their salaries. So Fig 8 shows the influence of WOM (the Adoption Fraction), we can find a positive influence from WOM on the adoption of electrical car.

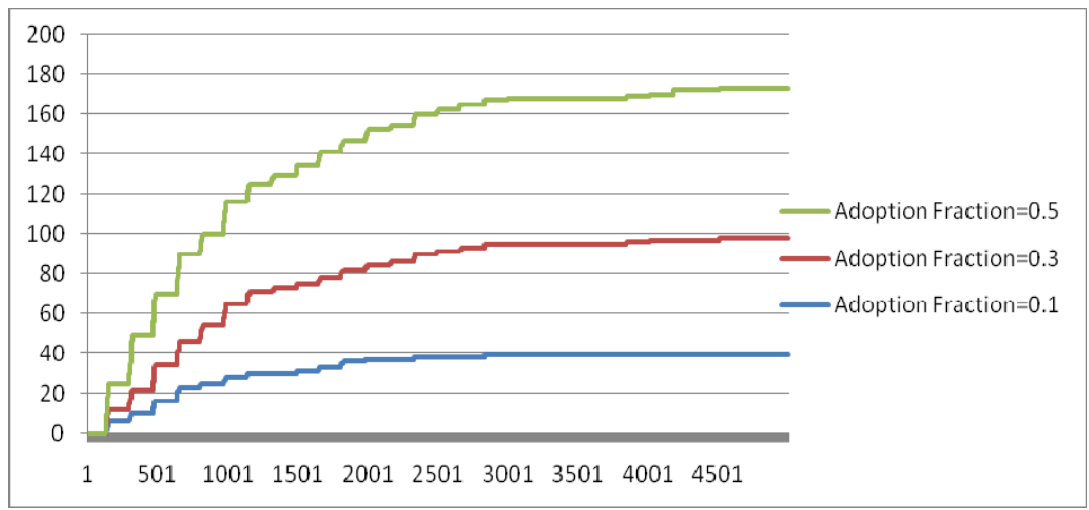

**Figure 8.** Electrical Car Distribution with the influence of WOM

## 5. Conclusion

In this paper we developed an agent-based simulation model to study the possible impact of different governmental interventions on the diffusion of such vehicles. Options that could be studied with our what-if analysis too include things like car parking charges, price of electrical car, energy awareness and word of mouth. In this paper we present a first case study related to the introduction of a new car park charging scheme at the University of Nottingham. We have developed an agent based model to simulate the impact of different car parking rates and other incentives on the uptake of electrical cars. The goal of this case study is to demonstrate the usefulness of agent-based modelling and simulation for such investigations.